\documentclass[usenatbib]{mn2e}
\usepackage[T1]{fontenc}
\usepackage{times}
\usepackage{lmodern}
\usepackage{graphicx}
\input{journaldef.sty}
\usepackage{txfonts}
\newcommand\UCHII{UCH\,{\sc ii}}
\newcommand\HII{H\,{\sc ii}}
\newcommand\CII{C\,{\sc ii}}

\newcommand\cmthree{cm$^{-3}~$}
\newcommand\emunit{pc~cm$^{-6}~$}

\newcommand\be{\begin{equation}}
\newcommand\ee{\end{equation}}
\newcommand\bea{\begin{eqnarray}}
\newcommand\eea{\end{eqnarray}}

\newcommand\ratioa{C76$\alpha$/C53$\alpha$}
\newcommand\ratiob{C92$\alpha$/C76$\alpha$}

\title[Modelling CRL observation towards W48A]{
Modelling Carbon Radio Recombination Line observation towards the Ultra-Compact \HII\ region W48A}

   \author[]{S. Jeyakumar$^{1,3}$\thanks{Email: sjk@astro.ugto.mx} and
          D. Anish Roshi$^{2}$\thanks{Email: aroshi@nrao.edu} \\
   $^1$ Departamento de Astronom{\'i}a, Universidad de Guanajuato, AP 144, Guanajuato CP 36000, Mexico \\
   $^2$ National Radio Astronomy Observatory\thanks{The National Radio Astronomy Observatory is
               a facility of the National Science Foundation operated under cooperative
               agreement by Associated Universities, Inc.}, Charlottesville, VA 22903-4608, USA \\
   $^3$ Raman Research Institute, C. V. Raman Avenue, Sadashivanagar, Bangalore - 560 080, India }

\begin{document}

  \date{}

  \pagerange{\pageref{firstpage}--\pageref{lastpage}} \pubyear{}

  \maketitle

\label{firstpage}

\begin{abstract}

    We model Carbon Recombination Line (CRL) emission 
    from the Photo Dissociation Region (PDR) surrounding  
    the Ultra-Compact (UC) \HII\ region W48A. Our modelling shows that the inner 
    regions ($A_V \sim 1$) of the
    \CII\ layer in the PDR contribute significantly to the CRL emission.
    The dependence of line ratios of CRL emission with the density of 
    the PDR and the far ultra-violet (FUV) 
    radiation incident on the region is explored over a
    large range of these parameters that are typical for the environments
    of \UCHII\ regions. We find that by observing a suitable set of CRLs it is
    possible to constrain the density of the PDR. 
    If the neutral density in the PDR is high ($\gtrsim 10^7$ \cmthree) CRL 
    emission is bright at high frequencies ($\gtrsim 20$ GHz), and absorption lines 
    from such regions can be detected at low frequencies ($\lesssim 10$ GHz). 
    Modelling CRL observations towards W48A shows  
    that the \UCHII\ region is embedded in a molecular cloud of density 
    of about $4~\times$ 10$^7$ \cmthree. 

\end{abstract}


   \maketitle

\section{Introduction}

HII regions with size less than about 0.1~pc are called Ultra Compact
HII regions (\UCHII\ regions). Their radio continuum emission is optically thick 
at frequencies below a few GHz indicating 
emission measures $\ga$10$^7$~\emunit and electron densities 
$\ga 10^4~$\cmthree \citep[cf.][]{wood&churchwell89a, kurtz.etal94, garay&lizano99}.
Molecular line observations reveal that \UCHII\ regions are embedded in dense molecular 
clouds \citep[e.g.][]{kim&koo03}. 
Line emission from many  high density tracer molecules, such as NH$_3$, 
and CS, is detected towards some of the \UCHII\ regions 
\citep[cf.][]{churchwell.etal90, garay&lizano99, kim&koo03}.
The \UCHII\ regions are bright at far-infrared wavelengths and modelling 
this emission suggests hot dust envelopes surrounding newly formed stars 
\citep{churchwell93,kurtz.etal94}.  
Some of these sources also show evidence for internal density and
velocity gradients within the ionised region 
\citep{sewilo.etal08, keto.etal2008, phillips.07}.
These observations imply that \UCHII\ regions are early evolutionary 
stages of massive stars, and embedded in dense massive molecular clouds 
with high optical extinctions \citep[cf.][]{churchwell02}.

The age of \UCHII\ regions estimated from the observed number
of such \HII\ regions in the Galaxy and the galactic star formation rate is 
$\sim$ 10$^5$ years. This age is found to be longer than 
the time scale of expansion of the ionised gas to a size of 
about 0.1 pc (dynamical age; a few times 10$^3$ years) 
at its sound speed \citep[cf.][]{wood&churchwell89b}. 
Many models have been proposed to resolve this inconsistency in age or 
otherwise called the {\it age paradox} 
\citep[cf.][and references therein]{franco.etal07, garay&lizano99}.
The proposed models include, (a) confinement of ionized gas due to thermal or turbulent pressure of
surrounding material, (b) confinement by ram pressure of infalling material or bow shocks
(c) champagne flows, (d) disk evaporation, (e) mass loaded stellar
wind.  Recently \citet{peters.etal2010a} proposed that `flickering' of 
the size of the ionized regions around massive stars due to 
shielding by dense filaments in accretion flow makes the size  
of the \HII\ region independent of the age of the star.
Many of these models need dense external medium surrounding the 
ionised gas \citep[e.g.][]{depree.etal95}. Indeed, observations of 
molecular and carbon recombination lines (CRL) have also shown the presence 
of high-density ($>$ 10$^5$~\cmthree) 
molecular material in the vicinity of \UCHII\ regions \citep{roshi.etal05b}.

The high-density molecular material surrounding 
\UCHII\ regions can be heated by the FUV radiation from the embedded 
massive stars, establishing a
Photo Dissociation Region (PDR) \cite[cf.][]{tielens&hollenbach85}. 
The PDR in the vicinity of the \HII\ region forms a 
thin layer, mostly consisting of ionised carbon (CII) and neutral hydrogen 
(HI). 
The physical conditions in this layer can be inferred by observing 
[CII], [OI] and ro-vibrational H$_2$ emission lines \cite[cf.][]{hollenbach&tielens97},
which are all in the infra-red band. The recombination of 
an electron with a carbon ion at high quantum numbers, n$\gtrsim$40, and
subsequent cascade produces CRLs in the radio
frequencies. Observations of these CRLs can also be used to infer the physical 
conditions of the PDR \citep[e.g.][] {roelfsema&goss92,natta.etal94, roshi.etal05a}. 
Amplification of the CRL by the background continuum emission 
makes it possible to detect these lines easily 
with the existing radio telescopes \citep[cf.][]{dupree&goldberg70}. Such stimulated CRL 
emission is observed towards many regions of ionised carbon in the Galaxy 
\citep[e.g.][and references therein]{natta.etal94,wyrowski.etal97,roshi.etal02}. 
\citet{roshi.etal05a} also detected stimulated CRL emission near 8.5 GHz 
from PDRs associated with a large number of \UCHII\ regions. 

Attempts have been made earlier to infer the physical conditions in the PDR
using CRL emission with the help of detailed PDR models \citep{gomez.etal98}. 
\citet{natta.etal94} studied the line intensity ratio of C91$\alpha$ (near 8.6 GHz) and C66$\alpha$ 
(near 24 GHz) transitions using a limited set of homogeneous PDR models 
with densities in the range 10$^4$ and 10$^7$~\cmthree and 
incident FUV field between 10$^3$ and 10$^6$~$G_0$, where
$G_0$ is the standard interstellar FUV field 
($1.6\times10^{-3}$ ergs cm$^{-2}$ s$^{-1}$) of \citet{habing68}.

However, based on the observational studies of the \UCHII\ regions
and their environment, the ranges of PDR and \UCHII\ region properties 
are expected to be larger. The densities of the molecular material near
\UCHII\ regions inferred from observations of high density tracer
molecule, NH$_3$, are in the range of a few times 10$^{4}$ to
10$^{8}$~\cmthree \citep[cf.][]{garay&lizano99}. The expected FUV
field at the surface of \UCHII\ regions due to the embedded O4 to B0
type stars ranges from 10$^{4}$ to greater than 10$^{7}~G_0$
\citep{wood&churchwell89b}.

Therefore, in this paper, we examine a complete set of PDR models, 
spanning a large range of densities and incident radiation field which are 
required to model the CRL emission observed towards \UCHII\ regions. 
The details of modelling are
discussed in Section~\ref{sec:pdr}. In Section~\ref{sec:crlmodel},
we show that a multitude of 
recombination transitions from PDRs near \UCHII\ regions 
can be used to constrain the physical conditions of this layer. 
We apply our model to interpret the observed CRL emission
towards W48A in Section~\ref{sec:compare}.

\section{PDR surrounding an \UCHII\ region}
\label{sec:pdr}

PDR is created at the interface between an \UCHII\ region and
the associated molecular cloud. During the early phases of the evolution of \HII\ regions, 
the existence of the PDR layer depends on the relative speed between the 
ionisation and the dissociation fronts \citep{bertoldi&draine96}. 
However, \UCHII\ regions can attain pressure equilibrium within the 
dense molecular cores \citep{kurtz.etal01}, and the high pressure cores 
can stop the expansion at a time scale of about $3\times10^4$ yr 
\citep{franco.etal00}. \citet{arthur.etal04} noted that the dust in 
the ionised region can stall the expansion even earlier. This time scale
is a factor of about 10 smaller than the age inferred from the observed
number of \UCHII\ regions \citep{wood&churchwell89b} 
suggesting that the \HII\ regions may be pressure confined
for a long time. CRL observations towards \UCHII\ regions 
also indicate that these \HII\ regions may be pressure confined
\citep{roshi.etal05a}. 

The PDRs can attain steady state if the age of these regions is much
larger than the time scales of cooling and molecular 
processes.  The cooling time scale is of the order of 
$4 \times (T/1000)^{-7/2}$~yr, 
which is much smaller than the 
evolutionary time scales \citep{hollenbach&natta95} for the
typical temperature of a few 10s to $\sim$ 1000 K in the PDR. 
Of the chemical processes, the slower hydrogen formation time scale, 
$\tau_{H_2}$, is about $5\times10^{8}/n$ yr, where $n$ is the 
hydrogen gas density in \cmthree\ \citep{goldsmidt&sternberg95}.  
For densities $\gtrsim$ 10$^{5}$~\cmthree, $\tau_{H_2}$ is $<$5000 yr, which
is shorter than the expected age or the time scale for pressure 
equilibrium. This suggests that the PDR surrounding the \UCHII\ 
regions can attain steady state in a relatively short time scale for
high ambient gas densities. 

For an expanding \UCHII\ region the PDR properties evolve with time
\citep{franco.etal90, roger&dewdney92}. 
In this case, the thermal equilibrium is achieved faster 
than the dynamical time scales \citep{hollenbach&natta95}, 
but the effect of time dependent chemistry might be important. 
In this paper, we consider the case where the PDR is in a steady state.

\subsection{Model for Carbon recombination lines}
\label{sec:model}
   \begin{figure}
   \centering
   \includegraphics[width=6cm]{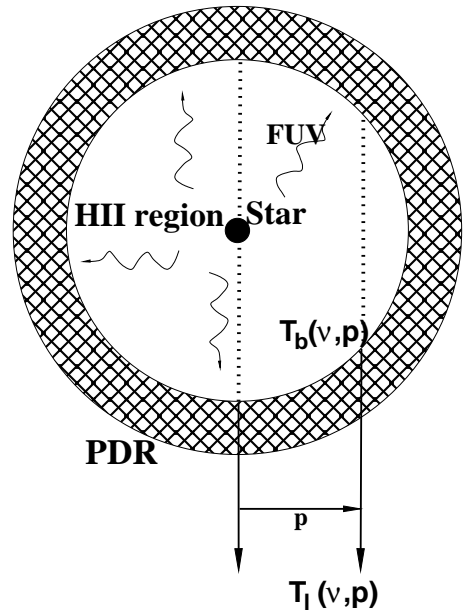}
   \caption{A schematic of the geometry used for the CRL intensity
calculations (not to scale). The PDR layer, shown as shaded region, surrounds
the \UCHII\ region. The FUV radiation from the central star in the \HII\ region
produces the PDR layer. The background continuum $T_b(\nu,p)$ for each ray 
with impact parameter, p, is calculated along the dotted lines. $T_l(\nu,p)$
is the calculated line brightness temperature corresponding to this ray.
    \label{sfig} }
    \end{figure}

We consider a spherical \UCHII\ region of radius, R$_{\rm HII}$, 
placed inside a molecular medium of density ${n}$, for modelling. 
The ionising star is assumed to be at the centre of the \UCHII\ region
and is stationary with respect to the molecular cloud.
The molecular material is heated by the FUV radiation from the
embedded star, establishing a PDR layer as shown schematically in Fig.~\ref{sfig}. 
The FUV field incident on the surface of the molecular medium, $G$, is 
specified in units of the mean interstellar radiation field $G_0$
\citep[cf.][]{lepetit.etal06}.
For a given FUV field and a molecular density, the PDR models solve the 
energy balance and the chemical network simultaneously. 
This solution provides the equilibrium gas temperature, T$_e$,
and the density of ionised carbon, n$_{\rm CII}$, as well as that of the electron, 
n$_{\rm e}$, as a function of depth into the molecular medium measured 
from the surface of incidence of FUV radiation \citep[cf.][]{lebourlot.etal93}. 
For the present work, these results for the PDR are obtained using the
Meudon-PDR code (version 13XI03) developed by \citet{lebourlot.etal93} 
where a 1~D, semi-infinite plane parallel slab of gas is heated
from one side by the FUV field. The standard values of all other parameters 
of this model, such as the heating and cooling processes, 
can be obtained from the above reference. 
We used the standard chemical network 
available with this code (file named Drcos.chi) which uses
a total of 69 chemical species and 539 chemical reactions.

For the calculation of CII level populations,
we approximate the PDR layer into N$_{slab}$ slabs each 
with a constant T$_{\rm e}$, n$_{\rm e}$ and n$_{\rm CII}$. 
The level populations of the CII levels are calculated for each slab.
The effects of deviation of the level populations 
from LTE on the line emission are characterised by the factors,  b$_n$ and $\beta_n$ 
\citep[cf.][]{dupree&goldberg70}. These factors are calculated using 
the code originally developed by \citet{salem&brocklehurst79} 
and later modified by 
\citet{walmsley&watson82} and \citet{payne.etal94}. 
The excitation of carbon levels are affected by the background radio continuum 
radiation and local gas collisions. 
Since the free-free optical depth of the PDR layer for the
frequencies of interest here is negligible, 
all the slabs receive the same background continuum intensity. 
The temperature of the 
background radiation incident at the 
surface of the PDR layer, at any given frequency is determined 
using the electron temperature and emission measure of the ionised gas in the 
spherical \UCHII\ region \citep{mezger&henderson67}.

For estimating the CRL intensities, we substituted the radial dependence 
of the physical quantities of the spherical PDR shell by the 1~D results 
obtained as described above. The emergent intensity is calculated for many 
rays ($\sim$ 200). Each ray is identified with an impact parameter, p 
(see Fig.~\ref{sfig}), which is measured from the centre of the sphere 
and has values in the range 0 to R$_{\rm HII}$.
The background continuum for each ray is calculated through the chord shown
as dashed lines in Fig~\ref{sfig}.

   \begin{figure}
	     \includegraphics[width=9cm,clip=]{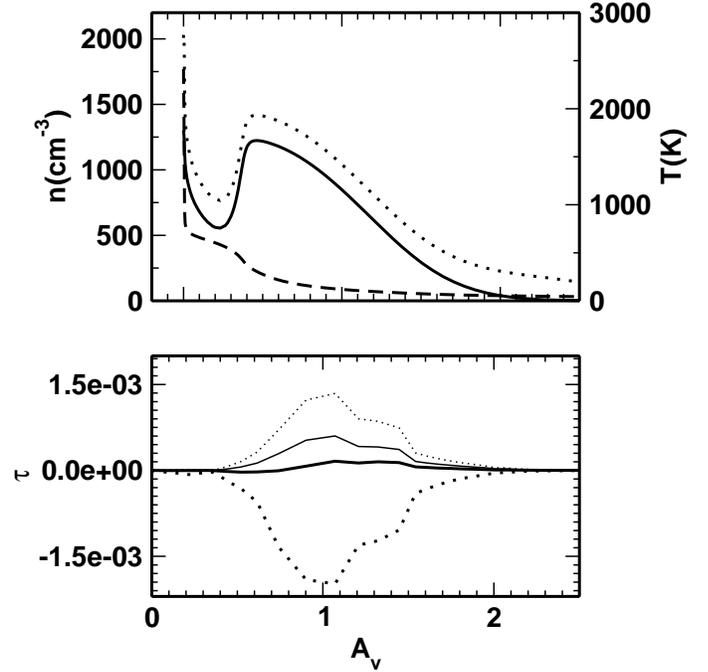}

     \caption{In the upper panel, the density of C$^+$ (solid) and 
electron (dotted) and the temperature of the gas (dashed) are plotted 
against the depth into the molecular cloud measured in $A_V$. 
These results are obtained using the PDR model, 
for a density of 10$^7$~\cmthree and an FUV field of 10$^4~G_0$.
In the lower panel, the CRL optical depths are plotted against $A_V$, 
for the same PDR properties shown in the upper panel.
The LTE optical depths for C53$\alpha$ and C76$\alpha$ are shown
by the thin solid and thin dashed lines respectively. Note that
the LTE optical depths have positive values for all $A_V$. The 
non-LTE optical depths of these transitions are shown with the
thick solid and thick dashed lines respectively.
The non-LTE optical depths have
been multiplied by a factor of 5 for displaying. 
    \label{pdrfig} }
   \end{figure}
%

\begin{figure}
\includegraphics[width=8.7cm,clip=]{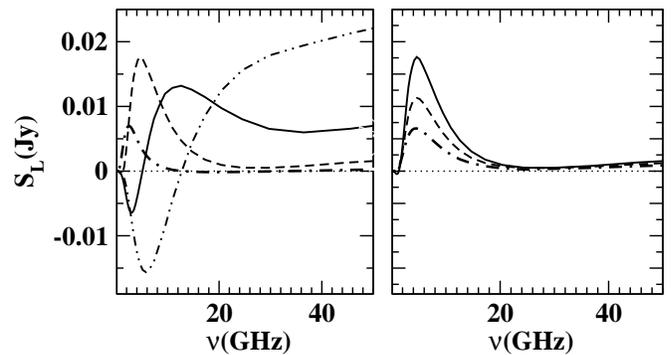}
\caption{The CRL flux density is plotted against 
frequency for different PDR densities (left) and
background FUV flux densities (right). In the left panel, 
the dot-dot-dashed, continuous, dashed and dot-dashed lines 
represent  line flux density for neutral densities 10$^8$, 10$^7$, 
10$^6$ and 10$^5$~\cmthree respectively. 
The FUV field for these calculations is kept at a constant value
of 10$^4$$G_0$.
In the right panel, the continuous, dashed and dot-dashed lines represent
$G$ of 10$^4$, 10$^5$ and 10$^6$$G_0$ respectively. The neutral
density used for these calculations is 10$^6$~\cmthree.
\label{ffig}}
\end{figure}

Following \citet{shaver75} the line radiation temperature 
for a ray with impact parameter, p, is given by, 
   \begin{eqnarray}
       T_l(\nu,p) &=& T_{l+c}(\nu,p) - T_c(\nu, p);
   \end{eqnarray}
where 
   \begin{eqnarray}
\nonumber    T_{l+c}(\nu,p) & = & T_b(\nu,p) exp{(-\tau^T_l-\tau^T_c)} \\
          & & + \int_0^{\tau^T_l+\tau^T_c} T_{ex}(\nu,\tau') exp{(-\tau')} d\tau', 
\\
\nonumber T_{c}(\nu,p) & = & T_b(\nu,p) exp{(-\tau^T_c)} \\
          & & + \int_0^{\tau^T_c} T_{ex}(\nu,\tau') exp{(-\tau')} d\tau' .
   \end{eqnarray}
In the above equations $\nu$ is the frequency of the transition.
$\tau'$ is the optical depth to a point in the PDR layer measured from
the outer boundary. $\tau^T$ is the total optical depth of the PDR layer. The
subscripts $l$ and $c$ denote the line and continuum respectively.
$T_b$ and $T_{ex}$ are the background radio continuum radiation temperature 
and excitation temperature of the CII levels respectively.
The surface averaged line radiation temperature is obtained as  
   \begin{displaymath}
        T_L(\nu) = { \int_0^{\rm R_{HII}} T_l(\nu, p)~2\pi~p~dp \over \int_0^{\rm R_{HII}}~2\pi~p~dp }.
   \end{displaymath}

\section{Formation of CRL in PDR}
\label{sec:crlmodel}

   \begin{figure*}
     \hbox{
     \includegraphics[width=8.5cm]{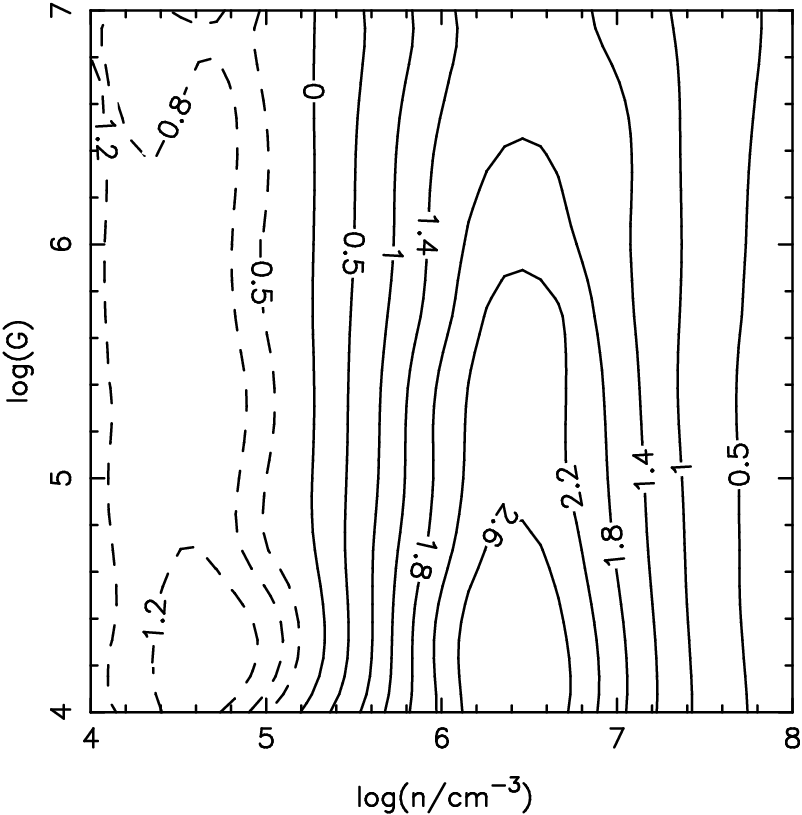}
     \includegraphics[width=8.5cm]{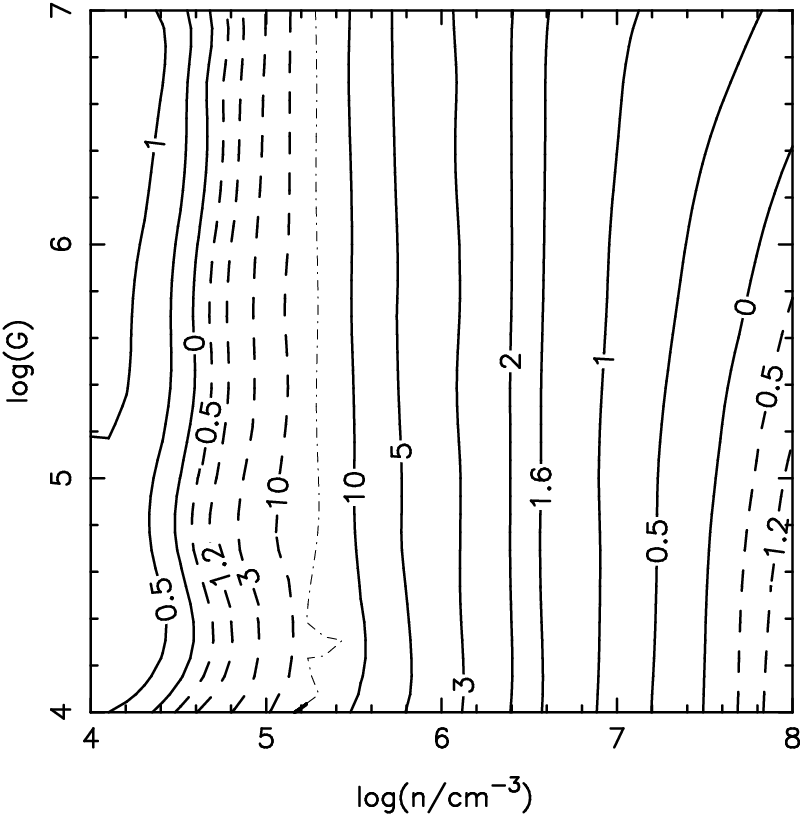}
      }
     \caption{
The CRL flux density ratios \ratioa (left panel) and \ratiob (right panel)
are plotted in the log(n/\cmthree)--log($G$) plane. 
The dashed and solid lines show contours of negative and positive ratios
respectively. The thin dot-dashed line indicates those ratios where the
denominator is zero. The CRL flux densities are obtained by considering
a \UCHII\ region with EM = $6.4\times10^7$~\emunit\ , T$_e$ = 9900~K and 
R$_{\rm HII}$ = 0.059~pc. The negative contours in both panels for densities
below $\sim$  10$^5$~\cmthree are due to the negative values of 
C76$\alpha$ (see Fig.~\ref{ffig}). In the right panel, the
C92$\alpha$ flux density is negative at very low densities ($< 10^{4.5}$~\cmthree) 
providing positive line ratio. The negative contours at very high densities 
($> 10^{7.5}$~\cmthree) are due to C92$\alpha$ flux density being negative.
     \label{lfig1} }
   \end{figure*}

For understanding the CRL formation in PDR, the contribution 
to the total CRL emission, of each thin shell within the PDR layer 
surrounding the \UCHII\ region is studied. 
The combined PDR and CRL model as described in Section~\ref{sec:pdr} is calculated 
for the observed parameters of W48A (see Section~\ref{sec:compare}). 
Using these calculations various characteristics of CRL formation can be understood.

The CRL optical depth as a function of depth into the PDR 
is obtained by considering a molecular cloud of density 10${^7}$ \cmthree\ heated 
by an FUV flux of 10$^4$ $G_0$. The 
densities of $C^+$ and electrons and the gas temperature provided by
the PDR model for these parameters  are plotted as a function of $A_V$
in Fig.~\ref{pdrfig} (upper panel). 
Using these quantities the LTE  and non-LTE optical depths of 
C53$\alpha$ (45.4764 GHz) and C76$\alpha$ (14.6973 GHz) are computed, 
which are also plotted as a function of
$A_V$ in Fig.~\ref{pdrfig} (lower panel). As seen in Fig.~\ref{pdrfig}
both LTE and non-LTE optical depths have large (absolute) values 
near $A_V$ $\sim$ 1 for the PDR properties considered for modelling.
The CRL optical depth is proportional to ${\rm EM}/{\rm T}_e^{{5/2}}$ 
and hence will have large values at those $A_V$ where this factor is maximum.
The non-LTE effect depends on a number of factors such as background
radiation field and electron and ion densities. It is expected that
the non-LTE optical depth becomes negative for a set of frequencies
resulting in partial masing of the recombination line (Shaver 1975). For example, the 
non-LTE optical depth calculated for the model parameters considered here indicates
that the intensity of C76$\alpha$ is amplified by stimulated emission
(see Fig.~\ref{pdrfig}).

From modelling the Spectral Eenergy Distributions (SEDs) of smaller HII regions 
(including the Hyper-Compact HII regions) over a wide range in frequency, \citet{keto.etal2008} 
infer density gradients,  with power-law indices between -1.5 and -2.5. 
The SEDs of \HII\ regions with such density gradient will be different from
those with uniform density, spherical ionized gas considered in the CRL modeling.
Typically such SED's have higher radio flux denisty near frequencies ($>$ 10 GHz) where
the uniform denisty models become optically thin \citep[e.g.][]{keto.etal2008}.
At frequencies less than about 15 GHz, stimulated emission of CRLs due to 
background radiation occurs and hence higher line flux density is expected 
compared to the uniform density models. 
Similarly if the lines at these frequencies are in absorption, once again large 
line flux densites are expected.
Therefore the model results presented here are aplicable for \UCHII\ regions 
that do not show evidence of density profiles in their SEDs. 

Such density gradients can also be present in the molecular gas 
in the vicinity of the HII regions. 
Since the thickness of the layer that contributes to the 
CRLs is very small, any variation of the physical parameters within this layer 
is not expected to contribute significantly for the above observed indices. 
Further, we compared the variation of the LTE optical depths inside a medium 
with a density gradient with a power-law index of -2.0 
(for $n=10^7$~\cmthree, $G= 10^5~G_0$ and $n=10^8$~\cmthree, $G= 10^6~G_0$),
with those of the appropriate constant density models such that the mean density 
of the CII layers in both the models are the same. The differences in the 
variation of the optical depth with distance are negligible between these models.
Therefore constant molecular gas density models are a good approximation 
for modelling the CRLs. 
However the densities inferred using the CRL emitting PDR layer
would imply higher densities at the surface of the HII region for
a power-law density medium as compared to a constant density medium.

\subsection{Dependence of CRL intensity on frequency}
\label{uch2pars}

The variation of CRL flux density with 
the frequency of the line transition for different PDR densities
(10$^5$, 10$^6$, 10$^7$ and 10$^8$~\cmthree) are shown in Fig.~\ref{ffig} (left panel).
An FUV field of $10^4~G_0$
and \UCHII\ region with parameters described above are used for the flux 
density calculations. Qualitatively, Fig.~\ref{ffig} (left panel) shows that, 
for a given FUV field, the maxima of line emission shifts to higher
frequencies with increasing density. 

The variation of CRL flux density for different line transition
with incident FUV field (10$^4$, 10$^5$ and 10$^6$ $G_0$) is 
shown in Fig.~\ref{ffig} (right panel). A neutral density 
of 10$^6$~\cmthree for the PDR and \UCHII\ region with parameters 
described above are used for these calculations. 
The figure shows that the flux density generally decreases with
increasing FUV field. This decrease in flux density is because
of the general increase in the PDR temperature with the FUV field. The CRL
optical depth inversely depends on the PDR temperature 
(${\rm T}_e^{-{5/2}}$) and hence the flux density decreases with the FUV field.

Examination of the frequency dependence of the CRL flux density 
(Fig.~\ref{ffig}) shows that for densities $\gtrsim$
10$^7$~\cmthree, it may be possible
to detect CRL in absorption at frequencies below $\sim$ 10 GHz. 
The absorption at low frequencies occurs since 
the level population of quantum states corresponding to these
frequencies approach the LTE value at higher densities.
The LTE excitation temperature for these transitions is smaller 
than the background radiation temperature of the
\UCHII\ region.

\subsection{CRL from an expanding UCHII region}
\label{subsec:expansion}

Modeling shows that, for densities typical for PDRs near \UCHII\ regions, it is possible
to detect line emission at smaller quantum numbers (ie
higher frequencies; $\gtrsim$ 20 GHz or so). This is evident from Fig.~\ref{ffig}
for densities $\gtrsim$ 10$^7$~\cmthree\ where line emission is detectable at
frequencies above $\sim$ 20 GHz. The non-LTE optical depth of these lines
are positive (see for example Fig.~\ref{pdrfig}) indicating that the line emission
is not dominated by stimulated emission.
Therefore, for the geometry of the PDR shown in Fig.~\ref{sfig}, detecting line emission
at frequencies $\gtrsim$ 20 GHz gives an unique opportunity to constrain the
expansion speed of the \UCHII\ region.
At these frequencies any expansion will produce a double profile
for the CRL emission and the separation between the two line
components gives a direct measure of the expansion of the \UCHII\ region. 
Application of this method to the \UCHII\ regions will be discussed further in a forthcoming paper.

\subsection{Dependence of CRL intensity on PDR parameters}
\label{subsec:pdrpars}

The ranges of PDR parameters over which we
present the model results are chosen based 
on the observational studies of the environments of the \HII\ regions.
The densities of the molecular material near \UCHII\ regions 
inferred from observations of high density tracer molecule, 
NH$_3$, are in the range of a few times 10$^{4}$ to  
10$^{8}$~\cmthree \citep[cf.][]{garay&lizano99}. The expected FUV 
field at the surface of \UCHII\ regions due to the embedded O4 to B0 
type stars ranges from 10$^{4}$ to greater than 10$^{7}~G_0$ 
\citep{wood&churchwell89b}. 
Thus we estimate the CRL intensities for molecular cloud densities 
ranging from 10$^4$ to 10$^8$~\cmthree and $G$ ranging from 
10$^4$ to 10$^7~G_0$. For the parameters of the \UCHII\ region, the 
observed values of W48A are used (see Section~\ref{sec:compare}).

The CRL emission is calculated for a set of 63 PDR models  
for a grid in the $G-n$ plane. The results are further re-gridded 
into a fine grid. 
In Fig~\ref{lfig1} we plot
the variation of the ratio of line intensities between the transitions
C92$\alpha$(8.3135 GHz), C76$\alpha$(14.6973 GHz) 
and C53$\alpha$(45.4764 GHz), in the $G-n$ plane.
The parameters which have been assumed for the FWHM of the telescope beam
and the distance to the cloud do not affect the line ratios.
The contours of the line ratio, \ratiob\,
clearly distinguish the density of the PDR but mostly independent
of the FUV field.
The line ratio, \ratioa\, shows  a similar dependence 
but two densities are possible for a given observed line
ratio. These figures show that observations at a suitable set of frequencies 
will be able to constrain the density of the PDR.

 \begin{figure}
     \includegraphics[width=8.5cm,clip=]{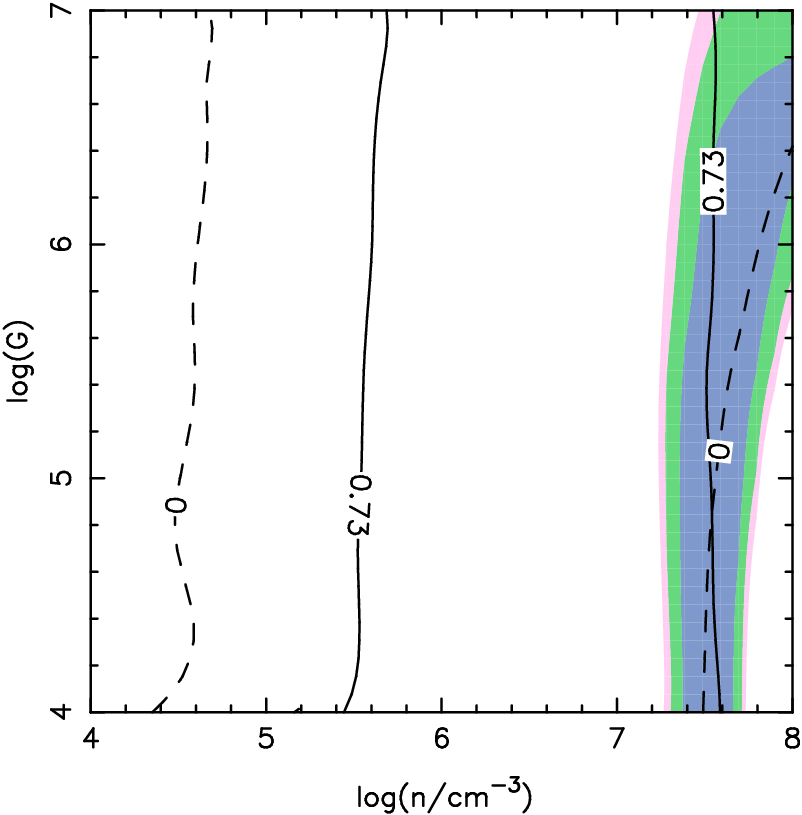}
     \caption{The contours of the observed line ratios \ratioa\ 
(solid) and \ratiob\ (dashed) towards W48A are plotted in the $G-n$ plane. 
The intersection of the contours of the observed CRL ratios gives a 
PDR density of about 4 $\times$ 10$^7$~\cmthree and an FUV field of about 7 $\times$ 10$^4 G0$.
The chi-square contours corresponding to 1(cyan), 2(green) and 3(magenta)  $\sigma$ 
confidence regions of the fit are plotted with decreasing levels of grey shaded regions.
     \label{cfig}
      }
   \end{figure}

\section{PDR surrounding the UCHII region W48A}
\label{sec:compare}

The \UCHII\ region W48A is located in the high mass star-forming region 
G35.20$-$1.74 at a distance of about 3.27 kpc \citep{wood&churchwell89a, zhang.etal09}.
W48A has been observed both in continuum and spectral line with the VLA at many 
frequencies by \citet{roshi.etal05b}. 
The continuum emission towards W48A can be well fit by a constant density spherical 
\UCHII\ region model. Based on this fit the estimated parameters of W48A are: the radius of the \UCHII\ region, $R_{\rm HII}$ is 0.059 pc;
the EM of the background continuum is $6.4\times10^7$~\emunit and the temperature of the ionised gas 
is 0.99$\times10^4$~K \citep{roshi.etal05b}. 
CRL emission is detected at 14 (C76$\alpha$) 
and 45 GHz (C53$\alpha$) and upper limits
have been provided at 8 (C92$\alpha$) and 4 GHz (C110$\alpha$, C111$\alpha$). 
At the distance of 3.27 kpc, the UCHII region subtends an 
angle of 3\arcsec.7, which is used to convert the model line temperature to line flux density.
For further details on the observations see \citet{roshi.etal05b}.

The observed value of \ratioa\ is 0.73$\pm$0.28 and that of \ratiob\ is $<0.3$
(after correcting for different beam sizes at different frequencies). 
Since C92$\alpha$ line flux density is an upper limit, it may
also represent an undetected  absorption line. 
Modelling also shows that the C92$\alpha$ transition can be in emission
or in absorption depending on the PDR density and background FUV field
(see Section~\ref{subsec:pdrpars}).
Therefore the ratio \ratiob\ is used as 0.0$\pm$0.3 to represent both positive and negative upper limits.
The CRL line ratio contours using the model results
are plotted in Fig~\ref{cfig} for those values observed towards W48A. 
The chi-square contours corresponding to 1(cyan), 2(green) and 3(magenta) $\sigma$ 
confidence regions of the fit are plotted with decreasing levels of grey shaded regions. 
The intersection of the contours of the observed CRL ratios gives a 
PDR density of about 4 $\times$ 10$^7$~\cmthree and an FUV field of about 7 $\times$ 10$^4 G0$.
The 1$\sigma$ confidence region, however, suggests that the allowed density and G values 
are larger due to large errors in the line ratios. Sensitive CRL observations are needed 
to further constrain the model parameters.  

The FUV field incident on the PDR, estimated from the parameters of 
the  stellar type (O8 -- O7.5) obtained by \citet{roshi.etal05b}, is $\lesssim 10^6 G0$
without considering any dust extinction for FUV photons. 
This value for G can be used to further constrain the of PDR density 
to a range between 2.5~$-$~7~$\times$ 10$^7$~\cmthree.  
The range of PDR densities obtained here is about 30\% smaller than the range 
obtained using a homogeneous slab model \citep{roshi.etal05b}.

\section{Conclusions}

We modelled the CRL emission from PDR layers surrounding an \UCHII\ region.
The depth dependence of temperature, ionised carbon and 
electron densities obtained from PDR models  have been incorporated into this
model. The non-LTE population of the carbon levels is calculated using
these temperatures and densities.
The CRL emission is presented over a range of PDR parameters. 
The results are shown in $G-n$ plane, where $G$ ranges from 10$^4$ to 10$^7$$G_0$
and the density ranges from 10$^4$ to 10$^8$~\cmthree.
Modelling the observed CRL emission towards W48A  
yields a density for the ambient medium of about $4\times10^7$~\cmthree.

\section*{Acknowledgements}
We are grateful to the anonymous referee for the critical
comments and suggestions. SJ thanks the support offered 
by the Raman Research Institute for a short visit.
We thank J. Le Bourlot for providing us the PDR code.  
This work was partially supported by PROMEP/103-5/07/2462 and
Conacyt CB-2009-01/130523 grants.

\bibliographystyle{mn2e}
\bibliography{uch2}

\begin{thebibliography}{41}
\expandafter\ifx\csname natexlab\endcsname\relax\def\natexlab#1{#1}\fi

\bibitem[{{Arthur} {et~al.}(2004){Arthur}, {Kurtz}, {Franco}, \& {Albarr{\'
  a}n}}]{arthur.etal04}
{Arthur} S.~J., {Kurtz} S.~E., {Franco} J., {Albarr{\' a}n} M.~Y., 2004, \apj,
  608, 282

\bibitem[{{Bertoldi} \& {Draine}(1996)}]{bertoldi&draine96}
{Bertoldi} F., {Draine} B.~T., 1996, \apj, 458, 222

\bibitem[{{Churchwell}(1993)}]{churchwell93}
{Churchwell} E., 1993, in ASP Conf. Ser. 35: Massive Stars: Their Lives in the
  Interstellar Medium, p.~35

\bibitem[{{Churchwell}(2002)}]{churchwell02}
---, 2002, \araa, 40, 27

\bibitem[{{Churchwell} {et~al.}(1990){Churchwell}, {Walmsley}, \&
  {Cesaroni}}]{churchwell.etal90}
{Churchwell} E., {Walmsley} C.~M., {Cesaroni} R., 1990, \aaps, 83, 119

\bibitem[{{de Pree} {et~al.}(1995){de Pree}, {Rodriguez}, \&
  {Goss}}]{depree.etal95}
{de Pree} C.~G., {Rodriguez} L.~F., {Goss} W.~M., 1995, Revista Mexicana de
  Astronomia y Astrofisica, 31, 39

\bibitem[{{Dupree} \& {Goldberg}(1970)}]{dupree&goldberg70}
{Dupree} A.~K., {Goldberg} L., 1970, \araa, 8, 231

\bibitem[{{Franco} {et~al.}(2007){Franco}, {Garc{\'{\i}}a-Segura}, {Kurtz}, \&
  {Arthur}}]{franco.etal07}
{Franco} J., {Garc{\'{\i}}a-Segura} G., {Kurtz} S.~E., {Arthur} S.~J., 2007,
  \apj, 660, 1296

\bibitem[{{Franco} {et~al.}(2000){Franco}, {Kurtz}, {Garc{\'{\i}}a-Segura}, \&
  {Hofner}}]{franco.etal00}
{Franco} J., {Kurtz} S.~E., {Garc{\'{\i}}a-Segura} G., {Hofner} P., 2000,
  \apss, 272, 169

\bibitem[{{Franco} {et~al.}(1990){Franco}, {Tenorio-Tagle}, \&
  {Bodenheimer}}]{franco.etal90}
{Franco} J., {Tenorio-Tagle} G., {Bodenheimer} P., 1990, \apj, 349, 126

\bibitem[{{Garay} \& {Lizano}(1999)}]{garay&lizano99}
{Garay} G., {Lizano} S., 1999, \pasp, 111, 1049

\bibitem[{{Goldshmidt} \& {Sternberg}(1995)}]{goldsmidt&sternberg95}
{Goldshmidt} O., {Sternberg} A., 1995, \apj, 439, 256

\bibitem[{{Gomez} {et~al.}(1998){Gomez}, {Lebron}, {Rodriguez}, {Garay},
  {Lizano}, {Escalante}, \& {Canto}}]{gomez.etal98}
{Gomez} Y., {Lebron} M., {Rodriguez} L.~F., {Garay} G., {Lizano} S.,
  {Escalante} V., {Canto} J., 1998, \apj, 503, 297

\bibitem[{{Habing}(1968)}]{habing68}
{Habing} H.~J., 1968, \bain, 19, 421

\bibitem[{{Hollenbach} \& {Natta}(1995)}]{hollenbach&natta95}
{Hollenbach} D., {Natta} A., 1995, \apj, 455, 133

\bibitem[{{Hollenbach} \& {Tielens}(1997)}]{hollenbach&tielens97}
{Hollenbach} D.~J., {Tielens} A.~G.~G.~M., 1997, \araa, 35, 179

\bibitem[{{Keto} {et~al.}(2008){Keto}, {Zhang}, \& {Kurtz}}]{keto.etal2008}
{Keto} E., {Zhang} Q., {Kurtz} S., 2008, \apj, 672, 423

\bibitem[{{Kim} \& {Koo}(2003)}]{kim&koo03}
{Kim} K., {Koo} B., 2003, \apj, 596, 362

\bibitem[{{Kurtz} {et~al.}(1994){Kurtz}, {Churchwell}, \&
  {Wood}}]{kurtz.etal94}
{Kurtz} S., {Churchwell} E., {Wood} D.~O.~S., 1994, \apjs, 91, 659

\bibitem[{{Kurtz} {et~al.}(2001){Kurtz}, {Franco}, {Garc{\'{\i}}a-Barreto},
  {Hofner}, {Garc{\'{\i}}a-Segura}, {de La Fuente}, \&
  {Esquivel}}]{kurtz.etal01}
{Kurtz} S., {Franco} J., {Garc{\'{\i}}a-Barreto} J.~A., {Hofner} P.,
  {Garc{\'{\i}}a-Segura} G., {de La Fuente} E., {Esquivel} A., 2001, in Revista
  Mexicana de Astronomia y Astrofisica Conference Series, pp. 45--48

\bibitem[{{Le Bourlot} {et~al.}(1993){Le Bourlot}, {Pineau Des Forets},
  {Roueff}, \& {Flower}}]{lebourlot.etal93}
{Le Bourlot} J., {Pineau Des Forets} G., {Roueff} E., {Flower} D.~R., 1993,
  \aap, 267, 233

\bibitem[{{Le Petit} {et~al.}(2006){Le Petit}, {Nehm{\'e}}, {Le Bourlot}, \&
  {Roueff}}]{lepetit.etal06}
{Le Petit} F., {Nehm{\'e}} C., {Le Bourlot} J., {Roueff} E., 2006, \apjs, 164,
  506

\bibitem[{{Mezger} \& {Henderson}(1967)}]{mezger&henderson67}
{Mezger} P.~G., {Henderson} A.~P., 1967, \apj, 147, 471

\bibitem[{{Natta} {et~al.}(1994){Natta}, {Walmsley}, \&
  {Tielens}}]{natta.etal94}
{Natta} A., {Walmsley} C.~M., {Tielens} A.~G.~G.~M., 1994, \apj, 428, 209

\bibitem[{{Payne} {et~al.}(1994){Payne}, {Anantharamaiah}, \&
  {Erickson}}]{payne.etal94}
{Payne} H.~E., {Anantharamaiah} K.~R., {Erickson} W.~C., 1994, \apj, 430, 690

\bibitem[{{Peters} {et~al.}(2010){Peters}, {Mac Low}, {Banerjee}, {Klessen}, \&
  {Dullemond}}]{peters.etal2010a}
{Peters} T., {Mac Low} M.-M., {Banerjee} R., {Klessen} R.~S., {Dullemond}
  C.~P., 2010, \apj, 719, 831

\bibitem[{{Phillips}(2007)}]{phillips.07}
{Phillips} J.~P., 2007, \mnras, 380, 369

\bibitem[{{Roelfsema} \& {Goss}(1992)}]{roelfsema&goss92}
{Roelfsema} P.~R., {Goss} W.~M., 1992, \aapr, 4, 161

\bibitem[{{Roger} \& {Dewdney}(1992)}]{roger&dewdney92}
{Roger} R.~S., {Dewdney} P.~E., 1992, \apj, 385, 536

\bibitem[{{Roshi} {et~al.}(2005{\natexlab{a}}){Roshi}, {Balser}, {Bania},
  {Goss}, \& {De Pree}}]{roshi.etal05a}
{Roshi} D.~A., {Balser} D.~S., {Bania} T.~M., {Goss} W.~M., {De Pree} C.~G.,
  2005{\natexlab{a}}, \apj, 625, 181

\bibitem[{{Roshi} {et~al.}(2005{\natexlab{b}}){Roshi}, {Goss},
  {Anantharamaiah}, \& {Jeyakumar}}]{roshi.etal05b}
{Roshi} D.~A., {Goss} W.~M., {Anantharamaiah} K.~R., {Jeyakumar} S.,
  2005{\natexlab{b}}, \apj, 626, 253

\bibitem[{{Roshi} {et~al.}(2002){Roshi}, {Kantharia}, \&
  {Anantharamaiah}}]{roshi.etal02}
{Roshi} D.~A., {Kantharia} N.~G., {Anantharamaiah} K.~R., 2002, \aap, 391, 1097

\bibitem[{{Salem} \& {Brocklehurst}(1979)}]{salem&brocklehurst79}
{Salem} M., {Brocklehurst} M., 1979, \apjs, 39, 633

\bibitem[{{Sewi{\l}o} {et~al.}(2008){Sewi{\l}o}, {Churchwell}, {Kurtz}, {Goss},
  \& {Hofner}}]{sewilo.etal08}
{Sewi{\l}o} M., {Churchwell} E., {Kurtz} S., {Goss} W.~M., {Hofner} P., 2008,
  \apj, 681, 350

\bibitem[{{Shaver}(1975)}]{shaver75}
{Shaver} P.~A., 1975, Pramana, 5, 1

\bibitem[{{Tielens} \& {Hollenbach}(1985)}]{tielens&hollenbach85}
{Tielens} A.~G.~G.~M., {Hollenbach} D., 1985, \apj, 291, 722

\bibitem[{{Walmsley} \& {Watson}(1982)}]{walmsley&watson82}
{Walmsley} C.~M., {Watson} W.~D., 1982, \apj, 260, 317

\bibitem[{{Wood} \& {Churchwell}(1989a)}]{wood&churchwell89a}
{Wood} D.~O.~S., {Churchwell} E., 1989a, \apj, 340, 265

\bibitem[{{Wood} \& {Churchwell}(1989b)}]{wood&churchwell89b}
---, 1989b, \apjs, 69, 831

\bibitem[{{Wyrowski} {et~al.}(1997){Wyrowski}, {Schilke}, {Hofner}, \&
  {Walmsley}}]{wyrowski.etal97}
{Wyrowski} F., {Schilke} P., {Hofner} P., {Walmsley} C.~M., 1997, \apjl, 487,
  L171

\bibitem[{{Zhang} {et~al.}(2009){Zhang}, {Zheng}, {Reid}, {Menten}, {Xu},
  {Moscadelli}, \& {Brunthaler}}]{zhang.etal09}
{Zhang} B., {Zheng} X.~W., {Reid} M.~J., {Menten} K.~M., {Xu} Y., {Moscadelli}
  L., {Brunthaler} A., 2009, \apj, 693, 419

\end{thebibliography}

\bsp

\label{lastpage}
\end{document}